\documentclass[onecolumn,prl]{revtex4}
\usepackage{bm}
\usepackage{epsfig}
\usepackage{amsmath}
\begin{document}
\title{Multi-channel scattering problems: Analytical approach for exact solution using Green's functions}
\author{Diwaker and Aniruddha Chakraborty\\
School of Basic Sciences, Indian Institute of Technology Mandi,\\
Mandi, Himachal Pradesh 175001, India.}
\date{\today }
\begin{abstract}
\noindent  We have proposed an analytical approach for exact solution of multi-channel scattering problems, in presence of Dirac Delta function couplings. Our solution is quite general and is valid for any set of potentials, if the Green's functions of the uncoupled potentials are known at the crossing point. Using our model, it is possible to express N-channel problem by N independent `single' channel problems and hence one can have a realistic solution of multi-channel scattering problem. In this paper we have shown that transition probability from first diabatic potential to any other potential can be easily evaluated using a very simple analytical formula - which only require value of eigenfunction of first uncoupled potential and value of Green's function of other uncoupled potential, at the crossing.
\end{abstract}
\maketitle

\section{Introduction}
\noindent Nonadiabatic transition due to potential curve crossing is one of the most possible mechanisms to effectively induce electronic transitions 
\cite{Naka1,Niki,Child,Osherov,Nikitin,Nakamura,Naka2,Shaik,Imanishi,Thiel,Yoshimori,Engleman,Mataga,Dev}. This is really an interdisciplinary topic and appears in various fields of physics, chemistry and biology \cite{Naka2,Nikitin,Niki,Child,Ani4}. The theory of non-adiabatic transitions dates back to $1932$, when the pioneering works were published by Landau \cite{Landau}, Zener \cite{Zener} and Stueckelberg \cite{Stueckelberg} and by Rosen and Zener
\cite{Rosen}. Since then there are numerous research papers in this area  \cite{Naka1,Niki,Child,Osherov,Nikitin,Nakamura,Naka2,Shaik,Imanishi,Thiel, Yoshimori, Engleman,Mataga,Dev}. For example, Osherov and Voronin solved the problem where two diabatic potentials are constant with exponential coupling
\cite{Voronin}. C. Zhu solved the problem where two diabatic potentials are exponential with exponential coupling \cite{Nikitinmodel}.
In our earlier publications we have reported analytical solution in those cases where two or more arbitrary potentials are coupled by Dirac Delta interactions \cite{Ani1,AniThesis,Ani2,AniBook,Ani3,Ani4,Ani5}. In reality, the transition between different diabatic potentials occurs most effectively at the crossing, because the necessary energy transfer between the electronic and nuclear degrees of freedom is minimum at that point. Therefore, it is important to analyse a model where coupling is localized in space rather than using a model where coupling is the same everywhere (i.e. constant coupling). Also this Dirac Delta function coupling model has the great advantage that it can be solved exactly \cite{Ani1,Ani2,Ani3,Ani4, Ani5, Ani6,AniBook,AniThesis}. Recently we have extended our research to deal with the cases where two potentials are coupled by any arbitrary interaction \cite{Ani6}. In this paper we consider the case of two or more arbitrary diabatic potentials with Dirac Delta couplings. The Dirac Delta coupling model has the advantage that it can be exactly solved, if the uncoupled diabatic potential has an exact solution. But the main problem is the fact that the calculation of the N-channel Green's function is impossible for a general system. In our most recent paper it is shown that if we use Dirac Delta coupling model, it is possible to express N-channel problem by N independent `single' channel problem and hence one can have a realistic solution of multi-channel scattering problem \cite{Ani7}. Using the same model it is shown that transition probability from first diabatic potential to any other potential can be easilty calculated by using single equation with two boundary conditions \cite{Ani7}. In this paper, using the same model, we have shown that the transition probability from first diabatic potential to any other potential can be easilty calculated using a very simple analytical formula - which only require value of eigenfunction of first uncoupled potential and value of Green's function of other uncoupled potential, at the crossing.

\section{Multi-Channel Scattering Problems}

\subsection{Two Channel Scattering Problems}

\noindent The time-independent Schr$\stackrel{..}{o}$dinger equation for any two state system is given by
\begin{equation}
\left(
\begin{array}{cc}
H_{11}(x) & V_{12}(x) \\
V_{21}(x) & H_{22}(x)
\end{array}
\right) \left(
\begin{array}{c}
\psi_1(x) \\
\psi_2(x)
\end{array}
\right) =E\left(
\begin{array}{c}
\psi_1(x) \\
\psi_2(x)
\end{array}
\right).
\end{equation}
\noindent The above two equations can be expressed as follows
\begin{eqnarray}
\psi_2(x)=V_{12}(x)^{-1}\left[E-H_{11}(x)\right]\psi_1(x)\;\;\; \text{and}\\\nonumber
\psi_2(x)=\left[E-H_{22}(x)\right]^{-1}V_{21}(x)\psi_1(x).
\end{eqnarray}
\noindent After eliminating $\psi_2(x)$ from the above two equations we get
\begin{equation}
\label{90}\left[E-H_{11}(x)\right]\psi_1(x) - V_{12}(x)\left[E-H_{22}(x)\right]^{-1}V_{21}(x)\psi
_1(x)=0.
\end{equation}
\noindent The above equations simplify considerably if $V_{12}(x)$ and $V_{21}(x)$ are
Dirac Delta function at $x_2$, which can be expressed using operator notation as $V=$
$K^{0}_{12}S=K_{12}^{0}$ $|x_2\rangle \langle x_2|$. The above equation now become
\begin{equation}
\left(\left[E-H_{11}(x)\right]-{K^{0}_{12}}^2|x_2\rangle \langle
x_2|\left[E-H_{22}\right]^{-1}|x_2\rangle \langle x_2|\right)\psi_1(x)=0.
\end{equation}
\noindent This may be written as
\begin{equation}
\left[H_{11}(x)+{K^{0}_{12}}^2\delta (x-x_2)G_2^0(x_2,x_2;E)\right]\psi_1(x)=E\psi
_1(x),
\end{equation}
\noindent where 
\begin{equation}
G_2^0(x_2,x_2;E)= \langle x_2|\left[E-H_{22}\right]^{-1}|x_2\rangle.
\end{equation}
Now the above equation can be written as
\begin{equation}
\left[H_{11}(x)+K_{12}^2\delta (x-x_2)\right]\psi_1(x)=E\psi_1(x),
\end{equation}
\noindent where $K_{12}^2= {K^{0}_{12}}^2 G_2^0(x_2,x_2;E)$. So the effect of second potential is entering into the above equation in terms of Dirac Delta function and $K_{12}$ in general is a complex valued function. If we express $H_{11}(x)$ in terms of kinetic energy and potential energy [$V_1(x)$] terms, the above equation becomes
\begin{equation}
-\frac{\hbar^2}{2 m} \frac{d^2\psi_1(x)}{dx^2}+ V_1(x)\psi_1(x)+K_{12}^2\delta(x-x_2)\psi_1(x)=E \psi_1(x).
\end{equation}
If we take the complex conjugate of the above equation we get
\begin{equation}
-\frac{\hbar^2}{2 m} \frac{d^2\psi^{*}_1(x)}{dx^2}+ V_1(x)\psi^{*}_1(x)+{K^{*}_{12}}^{2}\delta(x-x_2)\psi^{*}_1(x)=E \psi^{*}_1(x).
\end{equation}
Multiplying Eq.(8) from left by  $\psi^{*}_1(x)$ and Eq. (9) from left by  $\psi_1(x)$ we get
\begin{equation}
-\frac{\hbar^2}{2 m} \psi^{*}_1(x)\frac{d^2\psi_1(x)}{dx^2}+ V_1(x)\psi^{*}_1(x)\psi_1(x)+K_{12}^2\psi^{*}_1(x)\delta(x-x_2)\psi_1(x)=E \psi^{*}_1(x)\psi_1(x),
\end{equation}
and
\begin{equation}
-\frac{\hbar^2}{2 m} \psi_1(x)\frac{d^2\psi^{*}_1(x)}{dx^2}+ V_1(x)\psi_1(x)\psi^{*}_1(x)+{K^{*}_{12}}^2\psi_1(x)\delta(x-x_2)\psi^{*}_1(x)=E \psi_1(x)\psi^{*}_1(x).
\end{equation}
Now subtracting Eq. (10) from Eq. (9) we get
\begin{equation}
-\frac{\hbar^2}{2 m}\left[\psi^{*}_1(x)\frac{d^2\psi_1(x)}{dx^2} - \psi_1(x)\frac{d^2\psi^{*}_1(x)}{dx^2}\right]+\left({K_{12}}^2 - {K^{*}_{12}}^2 \right) \psi_1(x)\psi^{*}_1(x) \delta(x-x_2) =0.
\end{equation}
After simplification of Eq. (12), we get
\begin{equation}
-\frac{\hbar^2}{2 m} \frac{d}{dx}\left[\psi^{*}_1(x)\frac{d\psi_1(x)}{dx} - \psi_1(x)\frac{d\psi^{*}_1(x)}{dx}\right]+\left({K_{12}}^2 - {K^{*}_{12}}^2 \right) \psi_1(x)\psi^{*}_1(x) \delta(x-x_2) =0.
\end{equation}
Integrating the above equation from $x_2 - \epsilon $ to $x_2 + \epsilon $.
\begin{equation}
\frac{\hbar^2}{2 m}\left[\psi^{*}_1(x)\frac{d\psi_1(x)}{dx} - \psi_1(x)\frac{d\psi^{*}_1(x)}{dx}\right]^{x_2+\epsilon}_{x_2-\epsilon} = \left({K_{12}}^2 - {K^{*}_{12}}^2 \right) \psi_1(x_2)\psi^{*}_1(x_2).
\end{equation}
After simplification we get
\begin{equation}
\frac{\hbar}{2 m i}\left[\psi^{*}_1(x)\frac{d\psi_1(x)}{dx} - \psi_1(x)\frac{d\psi^{*}_1(x)}{dx}\right]^{x_2+\epsilon}_{x_2-\epsilon}
= \frac{2 {K^{0}_{12}}^2} {\hbar} Im \left[G_2^0(x_2,x_2;E)\right] \psi_1(x_2)\psi^{*}_1(x_2).
\end{equation}
So the transition probability from one diabatic potential to another is given by
\begin{equation}
T_{12}(E)= \frac{2 {K^{0}_{12}}^2} {\hbar} Im \left[G_2^0(x_2,x_2;E)\right] \psi_1(x_2)\psi^{*}_1(x_2).
\end{equation}
This is a very simple formula, for using $T_{12}(E)$ one only needs to know the value of $G_2^0(x,x;E)$ and $\psi_1(x)$ at $x=x_2$.

\subsection{Three Channel Scattering Problems}

\noindent The time-independent Schr$\stackrel{..}{o}$dinger equation for three state system is given by
\begin{equation}
\left(
\begin{array}{ccc}
H_{11}(x) & V_{12}(x) & V_{13}(x) \\
V_{21}(x) & H_{22}(x) & 0 \\
V_{31}(x) & 0 & H_{33}(x)
\end{array}
\right) \left(
\begin{array}{c}
\psi_1(x) \\
\psi_2(x) \\
\psi_3(x)
\end{array}
\right) =E\left(
\begin{array}{c}
\psi_1(x) \\
\psi_2(x) \\
\psi_3(x)
\end{array}
\right).
\end{equation}
\noindent This above matrix equation can be written in the following form
\begin{eqnarray}
\left[H_{11}(x)-E\right]\psi_1(x)+V_{12}(x)\psi_2(x)+V_{13}(x)\psi_3(x)=0,\\\nonumber
\left[H_{22}(x)-E\right]\psi_2(x)+V_{21}(x)\psi_1(x)=0\;\; \text{and} \\\nonumber
\left[H_{33}(x)-E\right]\psi_3(x)+V_{31}(x)\psi_1(x)=0.
\end{eqnarray}
\noindent The above equation after rearranging is given below
\begin{eqnarray}
\left[E-H_{11}(x)\right]\psi_1(x)-V_{12}(x)\psi_2(x)-V_{13}(x)\psi_3(x)=0,\\
\psi_2(x)=\left[E - H_{22}(x)\right]^{-1}V_{21}(x)\psi_1(x)\;\; \text{and} \\
\psi_3(x)= \left[E - H_{33}(x)\right]^{-1}V_{31}(x)\psi_1(x).
\end{eqnarray}
After eliminating both $\psi_2(x)$ and $\psi_3(x)$ from Eq. (19) we get
\begin{equation}
\left(H_{11}(x)+ V_{12}(x)\left[E - H_{22}(x)\right]^{-1} V_{21}(x) + V_{13}(x)\left[E - H_{33}(x)\right]^{-1}V_{31}(x)\right) \psi_1(x)= E \psi_1(x).
\end{equation}
The above equations are true for any general $V_{12}$, $V_{21}$, $V_{13}$ and $V_{31}$. 
The above equation simplify considerably if $V_{12}$, $V_{13}$, $V_{31}$ and $V_{21}$ are Dirac Delta functions,
which we write in operator notation as $V_{12}=V_{21}=K^{0}_{12} S=K^{0}_{12}|x_2\rangle \langle x_2|$ and 
$V_{13}=V_{31}=K^{0}_{13} S=K^{0}_{13}|x_3\rangle \langle x_3|$.
\noindent The above equation now becomes
\begin{equation}
\left(H_{11}(x)+{K^{0}_{12}}^2\delta(x-x_2)G_2^0(x_2,x_2;E)+{K^{0}_{13}}^2\delta (x-x_3)G_3^0(x_3,x_3;E)\right)\psi_1(x)= E \psi_1(x).
\end{equation}
\noindent After simplification we get
\begin{equation}
\left(H_{11}(x)+{K_{12}}^2\delta(x-x_2)+{K_{13}}^2\delta (x-x_3)\right)\psi_1(x)= E \psi_1(x),
\end{equation}
where ${K_{12}}^2={K^{0}_{12}}^2 G_2^0(x_2,x_2;E)$ and ${K_{13}}^2={K^{0}_{13}}^2 G_3^0(x_3,x_3;E)$. So the effect of second and third potential is entering into the above equation in terms of Dirac Delta functions. Now one can use a method similar to that used in last section, at $x_2$ to calculate the transition probability 
$\left[T_{12}(E)\right]$ from first potential to the second potential. The analytical expression for $ T_{12}(E)$ becomes
\begin{equation}
T_{12}(E)= \frac{2 {K^{0}_{12}}^2} {\hbar} Im \left[G_2^0(x_2,x_2;E)\right] \psi_1(x_2)\psi^{*}_1(x_2).
\end{equation}
Similarly transition probability  $\left[T_{13}(E)\right]$ from first potential to the third potential. The analytical expression for $ T_{13}(E)$ becomes
\begin{equation}
T_{13}(E)= \frac{2 {K^{0}_{12}}^2} {\hbar} Im \left[G_3^0(x_3,x_3;E)\right] \psi_1(x_3)\psi^{*}_1(x_3).
\end{equation}

\subsection{$N$ Channel Scattering Problems}

\noindent The time-independent Schr$\stackrel{..}{o}$dinger equation for
a $N$ state system, given by
\begin{equation}
\left(
\begin{array}{cccccc}
H_{11}(x) & V_{12}(x) & V_{13}(x)& . & . & V_{1N}(x) \\
V_{21}(x) & H_{22}(x) & 0 & 0 & 0 & 0 \\
V_{31}(x) & 0 & H_{33}(x) & 0 & 0 & 0 \\
. & . & . & . & . &  . \\
. & . & . & . & . &  . \\
V_{N1}(x) & 0 & 0 & .  & . & H_{NN}(x) \\
\end{array}
\right) \left(
\begin{array}{c}
\psi_1(x) \\
\psi_2(x) \\
\psi_3(x)
.\\
.\\
\psi_N(x)
\end{array}
\right) =E\left(
\begin{array}{c}
\psi_1(x) \\
\psi_2(x) \\
\psi_3(x) \\
.\\
.\\
\psi_N(x)
\end{array}
\right).
\end{equation}
\noindent The above matrix equation can be written in the following form
\begin{eqnarray}
\left[H_{11}(x)-E\right]\psi_1(x)+ \sum_{n=2}^{n=N} V_{1n}(x)\psi_n(x)=0\;\; \text{and} \\\nonumber
\left[H_{nn}(x)-E\right]\psi_n(x)+V_{n1}(x)\psi_1(x)=0.
\end{eqnarray}
\noindent The above equations after rearrangement are given below
\begin{eqnarray}
\left[E - H_{11}(x)\right]\psi_1(x)- \sum_{n=2}^{N} V_{1n}(x)\psi_n(x) = 0\;\; \text{and} \\
\psi_n(x)= \left[E - H_{nn}(x)\right]^{-1}V_{n1}(x)\psi_1(x)\;\; \text{for $n>1$}.
\end{eqnarray}
After eliminating $\psi_n(x)$ from Eq.(29) we get
\begin{eqnarray}
\left(H_{11}(x)+ \sum_{n=2}^{N} V_{1n}(x)\left[E - H_{nn}(x)\right]^{-1}V_{n1}(x)\right)\psi_1(x) = E \psi_1(x).
\end{eqnarray}
The above equations are true for any general $V_{1n}$ and $V_{n1}$. The above equation simplify considerably if $V_{1n}$ and  $V_{n1}$ are Dirac Delta functions, which we write in operator notation as $V_{1n}=V_{n1}=K^{0}_{1n} S=K^{0}_{1n} |x_n\rangle \langle x_n|$. The above equation now becomes
\begin{equation}
\left(H_{11}(x)+ \sum_{n=2}^{N}{K^{0}_{1n}}^2\delta (x-x_n)G_n^0(x_n,x_n;E)\right)\psi_1(x)= E \psi_1(x).
\end{equation}
\noindent After simplification we get
\begin{equation}
\left(H_{11}(x)+ \sum_{n=2}^{N}{K_{1n}}^2\delta(x-x_n)\right)\psi_1(x)= E \psi_1(x),
\end{equation}
where ${K_{1n}}^2={K^{0}_{1n}}^2 G_n^0(x_n,x_n;E)$. So the effect of all other potentials are entering into the above equation in terms of Dirac Delta functions.
Using our method, the transition probability from the first potential to the $n-th$ potential can be easily calculated. The analytical expression for $ T_{1n}(E)$ becomes
\begin{equation}
T_{1n}(E)= \frac{2 {K^{0}_{1n}}^2} {\hbar} Im \left[G_2^0(x_n,x_n;E)\right] \psi_1(x_n)\psi^{*}_1(x_n).
\end{equation}

\section{Conclusions:}

\noindent We have proposed a very simple general method for finding an exact analytical solution for the multi-channel quantum scattering problem in presence of a delta function couplings. Our solution is quite general and is valid for any potential. We have shown that the transition probability from first diabatic potential to any other potential can be easilty calculated using a very simple analytical formula - which only require value of eigenfunction of first uncoupled potential and value of Green's function of other uncoupled potential, at the crossing.

\section{Acknowledgments:}
\noindent A.C. would like to thank Prof. M. S. Child for his kind interest, suggestions and encouragements.

\end{document}